\documentclass[11pt]{article}
\usepackage{graphicx}
\usepackage{amssymb,amsmath,amsbsy}

\def\be{\begin{equation}}
\def\ee{\end{equation}}
\def\ba{\begin{eqnarray}}
\def\ea{\end{eqnarray}}

\begin{document} 

\title{A Two-Higgs Doublet Model With Remarkable CP Properties
}

\author{P.\ M.\ Ferreira$^{a,b}$ and Jo\~{a}o P.\ Silva$^{a,c}$\\
$^a$ Instituto Superior de Engenharia de Lisboa,\\
    Rua Conselheiro Em\'{\i}dio Navarro,
    1900 Lisboa, Portugal \\
$^b$ Centro de F\'{\i}sica Te\'{o}rica e Computacional,\\
        Universidade de Lisboa,
    Av.\ Prof.\ Gama Pinto 2,
    1649-003 Lisboa, Portugal\\
$^c$ Centro de F\'{\i}sica Te\'{o}rica de Part\'{\i}culas,\\
    Instituto Superior T\'{e}cnico,
    P-1049-001 Lisboa, Portugal}

\date{\today}

\maketitle \noindent {\bf Abstract.}
We analyze generalized CP symmetries of two-Higgs doublet
models, extending them from the scalar to the fermion sector of
the theory. We show that, with a single exception, those symmetries
imply massless fermions. The single model which accommodates a
fermionic mass spectrum compatible with experimental data possesses a
remarkable feature. It displays a new type of spontaneous CP violation,
which occurs not in the scalar sector
responsible for the symmetry breaking mechanism but, rather,
in the fermion sector.

\section{\label{sec:intro}Introduction}

Although the Standard Model (SM) of electroweak interactions
has had extraordinary experimental confirmation,
the Higgs mechanism of the symmetry breaking responsible for
giving mass to the particles remains largely untested.
In particular, there is no fundamental reason why the SM should
have only one Higgs. In fact, one of the simplest extensions
of the SM is the Two-Higgs Doublet Model (THDM) proposed by
Lee~\cite{Lee}, where there are five scalar particles (three
neutral and one charged). One of the main features of this model, and
the main reason Lee proposed it,
is that it allows for the possibility of the vacuum of the theory
spontaneously breaking the CP symmetry .

The most general scalar potential
in the THDM contains 14 real parameters, as opposed to
the 2 in the SM potential. There is thus a great interest in
reducing the number of free parameters in the THDM, through the
imposition of some symmetry in the theory. One such class of
symmetries are the so-called generalized CP (GCP) 
transformations - field transformations which combine CP with
unitary transformations. As far as we know,
GCP were first discussed in \cite{GCP1}. 
Their explicit use for quarks appeared in
\cite{GCP2} and
GCP in the scalar sector was initially developed
by the Viena group in \cite{GCP3,GCP4,GCP_NFC}.
GCP symmetries are quite interesting
in that they reduce immensely to number of free parameters in the potential.
Recently \cite{Ivanov1,FHS} it was
shown that, from the  point of view of the scalar potential,
these GCP symmetries fall unto three categories, designated
in~\cite{FHS} by CP1, CP2 and CP3. The CP1 class corresponds to the
standard CP transformation. The CP3 models, unlike CP1 and CP2,
correspond to a continuous symmetry.

In this paper we will show that CP2 cannot be extended from the scalar
sector to the fermionic one without forcing at least one fermion
to have zero mass. Likewise, it will be shown that, out of the infinite
number of possible CP3 models, only one can be extended to the fermionic
sector without inducing zero fermion masses. Thus we are able to
drastically curtail the number of models which includes an
acceptable fermion sector.

Having found only one acceptable GCP THDM, we will then proceed to
analyze its CP properties. We conclude that the model has a new type of
spontaneous CP breaking. In this model, CP violation does not emerge from
explicit CP breaking in the Yukawa sector, as it does in the SM; nor does
it emerge from spontaneous CP violation in the scalar sector, induced by
the vacuum of the model, as in the Lee model. Rather, in the model herein
presented, the lagrangian is
explicitly CP conserving and the scalar sector preserves CP, even after
spontaneous symmetry breaking. Nonetheless, the vacuum expectation values
generated by the scalar sector induce a CP violating phase in the quark
sector.
As far as we know, this is the first illustration in the literature of
this peculiar type of spontaneous CP violation.

This paper is structured
as follows: in section~\ref{sec:scalar} we present the scalar potential
of the theory and review recent results on its possible symmetries, as
well as the impact that GCP has on its parameters. In section~\ref{sec:yuk}
we show how GCP can be extended to the fermion sector, and we prove that,
with a single exception, GCP always implies massless fermions; we also
perform a fast fit to experimental constraints.
In  section~\ref{sec:spontaneous} we discuss the CP properties of this model,
and show that it has a new type of spontaneous CP breaking.
We present our conclusions in section~\ref{sec:conclusions}.

\section{\label{sec:scalar}The scalar potential}

Let us first review the scalar sector of the THDM, and what
is known of its symmetries.  In the THDM one introduces
two hypercharge $Y = 1/2$ Higgs
scalar doublets. The most general THDM scalar potential which
is renormalizable and compatible with the gauge symmetries of the SM
is given by
\begin{eqnarray}
V_H
&=&
m_{11}^2 \Phi_1^\dagger \Phi_1 + m_{22}^2 \Phi_2^\dagger \Phi_2
- \left[ m_{12}^2 \Phi_1^\dagger \Phi_2 + \textrm{H.c.} \right]
+
\tfrac{1}{2} \lambda_1 (\Phi_1^\dagger\Phi_1)^2
+ \tfrac{1}{2} \lambda_2 (\Phi_2^\dagger\Phi_2)^2
\nonumber\\[6pt]
&+&
\lambda_3 (\Phi_1^\dagger\Phi_1) (\Phi_2^\dagger\Phi_2)
+ \lambda_4 (\Phi_1^\dagger\Phi_2) (\Phi_2^\dagger\Phi_1)
+
\left[
\tfrac{1}{2} \lambda_5 (\Phi_1^\dagger\Phi_2)^2
+ \lambda_6 (\Phi_1^\dagger\Phi_1) (\Phi_1^\dagger\Phi_2)
\right.
\nonumber\\[6pt]
&+&
\left.
\lambda_7 (\Phi_2^\dagger\Phi_2) (\Phi_1^\dagger\Phi_2)
+ \textrm{H.c.}
\right],
\label{VH1}
\end{eqnarray}
and it involves 14 parameters.
Here $m_{11}^2$, $m_{22}^2$, and $\lambda_1,\cdots,\lambda_4$
are real parameters, while, in general,
$m_{12}^2$, $\lambda_5$, $\lambda_6$ and $\lambda_7$
are complex. ``H.c.''~stands for Hermitian conjugation.

This large number of  parameters may be reduced by requiring that
the model be invariant for some type of symmetry. Namely, we may require
invariance for field transformations of the form
\be
\Phi_a \rightarrow S_{ab} \Phi_b,
\ee
for an $S$ matrix belonging to $U(2)$, which are known as
Higgs family symmetries. Another possibility~\footnote{Both field
transformations discussed here obey a basic requirement, namely leaving the
kinetic terms of the lagrangian unaffected.} consists on the
so-called generalized CP symmetries \cite{GCP1,GCP2,GCP3,GCP4,GCP_NFC}, requiring invariance under
the following field transformations:
\be
\Phi_a \rightarrow X_{ab} \Phi_b^\ast,
\label{eq:GCP}
\ee
with $X \in U(2)$ as well.
Each of these symmetries corresponds to a different model, with different
phenomenology. One may wonder {\em how many} different choices of
scalar potentials can one make in the THDM.
Recently it has been shown that applying the
symmetries with any possible choices for $S$ and $X$ leads only to six
classes of scalar potentials \cite{Ivanov1,FHS}.
The impact of the six classes on the parameters are
given in Table~\ref{master1}, for specific basis choices.
%
\begin{table*}[ht!]
\caption{Impact of the symmetries on the coefficients
of the Higgs potential in a specific basis.
See Ref.~\cite{FHS} for more details.}
\begin{tabular}{ccccccccccc}
\hline \hline
symmetry & $m_{11}^2$ & $m_{22}^2$ & $m_{12}^2$ &
$\lambda_1$ & $\lambda_2$ & $\lambda_3$ & $\lambda_4$ &
$\lambda_5$ & $\lambda_6$ & $\lambda_7$ \\
\hline
$Z_2$ &   &   & 0 &
   &  &  &  &
   & 0 & 0 \\
$U(1)$ &  &  & 0 &
 &  & &  &
0 & 0 & 0 \\
$U(2)$ &  & $ m_{11}^2$ & 0 &
   & $\lambda_1$ &  & $\lambda_1 - \lambda_3$ &
0 & 0 & 0 \\
\hline
CP1 &  &  & real &
 & &  &  &
real & real & real \\
CP2 &  & $m_{11}^2$ & 0 &
  & $\lambda_1$ &  &  &
   &  & $- \lambda_6$ \\
CP3 &  & $m_{11}^2$ & 0 &
   & $\lambda_1$ &  &  &
$\lambda_1 - \lambda_3 - \lambda_4$ (real) & 0 & 0 \\
\hline \hline
\end{tabular}
\label{master1}
\end{table*}
%
This result is already extremely important: there are only six possible
different types of physical models in the scalar sector.
The first three may be obtained from family symmetries alone,
and are not the subject of this work.
The next three may be obtained from GCP symmetries alone, and we
will study them in detail.
Let us begin by explaining what the
difference is between them. Since
any basis of fields $\{\Phi_1 , \Phi_2\}$ has to give the same physical
results, we will use that liberty of basis choice to simplify the GCP field transformations of
Eq.~\eqref{eq:GCP}. In fact it has been shown
that there is always a basis of scalar fields, for which
the most general GCP transformation matrix $X$ may be brought
to the form
\be
X_\theta =
\left[
\begin{array}{cc}
c_\theta & s_\theta \\
- s_\theta &  c_\theta
\end{array}
\right],
\label{simple_GCP}
\ee
where $0 \leq \theta \leq \pi/2$ \cite{simpleGCP}.
Henceforth,
$c = \cos$, $s = \sin$ and the subindices indicate the angle.
The classes CP1, CP2, and CP3,
arise respectively from $X_0$, $X_{\pi/2}$, and $X_\theta$
for any one (or several) $\theta$ excluding $0$ and $\pi/2$.
CP1 is the usual CP symmetry. CP2, like CP1, is a discrete
symmetry, although it eliminates far more parameters in the
potential than CP1 does. CP3 corresponds to an infinite
number of possibilities, since the angle $\theta$ varies in a
continuous interval. We will now analyze the extension of
these symmetries to the fermion sector. Although we will only
look at the quark sector, the same conclusions can be trivially
extended to leptons.

\section{\label{sec:yuk}Yukawa interactions}

In the previous section we have seen that there are only three classes
of models with GCP symmetries. However, that study considered only
the scalar sector, not the whole lagrangian. We will now show how
GCP can be extended to the fermion sector, and what consequences arise
thereof. The scalar-quark Yukawa interactions of the THDM may be written as
\be
- {\cal L}_Y = \bar{q}_L (\Gamma_1 \Phi_1 + \Gamma_2 \Phi_2) n_R +
\bar{q}_L (\Delta_1 \tilde{\Phi}_1 + \Delta_2 \tilde{\Phi}_2) p_R + \textrm{H.c.},
\ee
where $q_L = (p_L, n_L)^\top$ ($n_R$ and $p_R$) is a vector in the 3-dimensional
generation space of left-handed doublets
(right-handed charge $-1/3$ and $+2/3$) quarks.
$\Gamma_1$, $\Gamma_2$, $\Delta_1$, and $\Delta_2$ are $3 \times 3$ matrices.

We now wish to impose GCP on scalars and also
on fermions. For the quark fields, the GCP transformations
take the form
\ba
q_L &\rightarrow X_\alpha  \gamma^0 C q^\ast_L,\nonumber \\
n_R &\rightarrow X_\beta  \gamma^0 C n^\ast_R,\nonumber \\
p_R &\rightarrow X_\gamma  \gamma^0 C p^\ast_R,
\ea
where $\gamma^0$ ($C$) is the Dirac (charge-conjugation) matrix.
As was the case with the scalar GCP transformations, we can also
simplify the above with an appropriate basis choice for the fermion
fields. In fact, we may take $X$ to the simple form \cite{simpleGCP}
\be
X_\alpha =
\left[
\begin{array}{ccc}
c_\alpha & s_\alpha & 0\\
- s_\alpha &  c_\alpha & 0\\
0 & 0 & 1
\end{array}
\right],
\label{simple_GCP_2}
\ee
where $0 \leq \alpha \leq \pi/2$,
and similarly for $X_\beta$ and $X_\gamma$.
GCP is a good symmetry of ${\cal L}_Y$ if and only if
$\Gamma_b^\ast = (X_\theta)_{ab} X_\alpha^\dagger \Gamma_a X_\beta$,
which we may write as
\ba
X_\alpha \Gamma_1^\ast - (c_\theta \Gamma_1 - s_\theta \Gamma_2) X_\beta = 0,
\nonumber\\
X_\alpha \Gamma_2^\ast - (s_\theta \Gamma_1 + c_\theta \Gamma_2) X_\beta = 0.
\label{geneq}
\ea
Since they are complex,
this gives us 36 equations in the 36 unknown real and imaginary
parts of the various entries of the $\Gamma$ matrices.
Because of Eq.~(\ref{simple_GCP_2}),
they break into four blocks,
which we designate by $mn$,
$m3$,
$3n$,
and $33$,
where $m$ and $n$ can take the values 1 or 2.
In each block we have a system of homogeneous linear equations;
the parameters are zero unless the determinant of the system
vanishes.
For instance, the real (Re) and imaginary (Im) parts of $(\Gamma_1)_{33}$ and
$(\Gamma_2)_{33}$ in Eq.~(\ref{geneq}) give the following set of
linear equations:
\begin{eqnarray}
\left[
\begin{array}{cc}
1 - c_\theta & s_\theta\\
-s_\theta & 1 - c_\theta
\end{array}
\right]\,
\left[
\begin{array}{c}
\mbox{Re}\left(\Gamma_1\right)_{33} \\
\mbox{Re}\left(\Gamma_2\right)_{33}
\end{array}
\right] = 0,
\nonumber \\
\left[
\begin{array}{cc}
1 + c_\theta & -s_\theta\\
s_\theta & 1 + c_\theta
\end{array}
\right]\,
\left[
\begin{array}{c}
\mbox{Im}\left(\Gamma_1\right)_{33} \\
\mbox{Im}\left(\Gamma_2\right)_{33}
\end{array}
\right] = 0.
\label{sist2}
\end{eqnarray}
The determinant of the first of these matrices gives $2(1-c_\theta)$,
whereas the second one is equal to $2(1+c_\theta)$.
Because of the
limited range of variation of $\theta$,
the second determinant is never zero - which
means that the $\Gamma_{33}$ coefficients will always be real.
As for the first of these determinants,
it is only zero if $\theta = 0$ (the standard CP definition).
For any other case, the $33$ entries of the $\Gamma$ matrices will be zero.

A similar reasoning applies to the $m3$ entries.
Eq.~(\ref{geneq}) gives us a set of $4\times 4$ equations
on the real and imaginary parts of the Yukawa couplings,
with corresponding determinants given by
$4(c_\alpha - c_\theta)^2$ and $4(c_\alpha + c_\theta)^2$,
respectively.
Since $\alpha$ has the same range of variation than $\theta$,
these determinants only vanish in very specific situations;
they vanish when
$\alpha = \theta$ or $\alpha = \theta = \pi/2$,
respectively.
In the same manner,
the $3n$ equations give rise to determinants equal
to $4(c_\beta - c_\theta)^2$ and $4(c_\beta + c_\theta)^2$ for the
real and imaginary parts,
respectively,
with analogous conditions for vanishing determinants.

Finally,
the $mn$ block from Eq.~(\ref{geneq}) gives us a set of
$8\times 8$ equations for the real and imaginary parts of the corresponding
Yukawas.
The corresponding determinants are
\begin{equation}
16\, (c_\theta \pm c_{\alpha + \beta})^2\,  (c_\theta \pm c_{\alpha - \beta})^2,
\label{mndet}
\end{equation}
with the ``+'' signs corresponding to the equations for the imaginary
parts.
We performed a thorough analysis of the conditions for vanishing
determinants,
and the results are summarized in Table~\ref{master2}.
%
\begin{table}[ht!]
\caption{Impact of the GCP symmetry on $\Gamma_1$ and $\Gamma_2$.
We separate real (Re) and imaginary (Im) components.}
\begin{center}
\begin{tabular}{ccc}
\hline \hline
$\Gamma_a$ matrix & component & condition for \\
element &  & vanishing determinant \\
\hline
33 & Im  &  impossible\\
   &  Re &  $\theta=0$\\
\hline
13, 23 & Im & $\alpha=\theta=\pi/2$\\
       & Re & $\alpha = \theta$\\
\hline
31, 32 & Im & $\beta=\theta=\pi/2$\\
       & Re & $\beta = \theta$\\
\hline
11, 12, 21, 22 & Im & $\theta = \pi - \alpha - \beta$\\
               & Re & $\theta = \alpha + \beta$ or $\theta = \alpha - \beta$\\
               &    &   or $\theta = \beta - \alpha$\\
               \hline \hline
\end{tabular}
\end{center}
\label{master2}
\end{table}
%
Similar results hold for the charged $+2/3$ quark
matrices $\Delta_1$ and $\Delta_2$, with $\beta \rightarrow \gamma$.

After spontaneous symmetry breaking,
the fields acquire the vacuum expectation values (vevs)
$v_1/\sqrt{2}$ and $v_2 e^{i \delta}/\sqrt{2}$;
where $v_1$ and $v_2$ are real,
without loss of generality.
It is convenient to rotate into
the so-called Higgs basis
$\{H_1, H_2 \}$ through $\Phi_a = U_{ab} H_b$,
where
\be
U^\dagger
=
\frac{1}{v}
\left[
\begin{array}{cc}
v_1 & v_2 e^{-i \delta}\\
v_2 & - v_1 e^{-i \delta}
\end{array}
\right],
\label{eq:HBT}
\ee
is unitary, and $v = \sqrt{v_1^2 + v_2^2} = (\sqrt{2} G_F)^{-1/2}$.
This rotates the vev into $H_1$ allowing us
to parametrize
\be
H_1
=
\left[
\begin{array}{c}
G^+\\
(v + H^0 + i G^0)/\sqrt{2}
\end{array}
\right],
\ \ \
H_2
=
\left[
\begin{array}{c}
H^+\\
(R + i I)/\sqrt{2}
\end{array}
\right],
\ee
where $ G^+ $ and $ G^0 $ are the Goldstone bosons,
which,
in the unitary gauge,
become the longitudinal components of the $ W^+ $ and of the $ Z^0 $,
and $ H^0 $, $ R $ and $ I $ are real neutral fields.
In the new scalar basis, the Yukawa coupling matrices become
$\Gamma_a^{H} = \Gamma_b U_{ba}$,
$\Delta_a^{H} = \Delta_b U_{ba}^\ast$,
%
%
and the scalar couplings are also rotated.
Finally,
the fermion mass basis is obtained through transformations
$U_\alpha$ with $\alpha = d_L, d_R, u_L, u_R$ that
diagonalize the quark couplings to $H_1$,
\ba
(v/\sqrt{2})\ U_{d_L}^\dagger \Gamma_1^{H} U_{d_R}
& = &
D_d = \textrm{diag} (m_d, m_s, m_b)\ ,
\nonumber\\
(v/\sqrt{2})\ U_{u_L}^\dagger \Delta_1^{H} U_{u_R}
& = &
D_u = \textrm{diag} (m_u, m_c, m_t )\ ,
\ea
while the couplings to $H_2$ become (we follow the
notation of \cite{details1,BS})
\ba
(v/\sqrt{2})\ U_{d_L}^\dagger \Gamma_2^{H} U_{d_R}
& = &
N_d\ ,
\nonumber\\
(v/\sqrt{2})\ U_{u_L}^\dagger \Delta_2^{H} U_{u_R}
& = &
N_u\ .
\label{eq:N}
\ea
The Yukawa lagrangian may then be written as
\ba
- \frac{v}{\sqrt{2}} {\cal L}_Y
&=&
({\bar u}_L V,\ {\bar d}_L )
( D_d  H_1 + N_d H_2 )\ d_R
\nonumber\\
&+&
( {\bar u}_L,\ {\bar d}_L V^\dagger )
( D_u  \tilde{H}_1 + N_u  \tilde{H}_2 )\ u_R + H.c.\ ,
\label{eq:yukhiggs}
\nonumber\\
\ea
where $V = U_{u_L}^\dagger U_{d_L}$ is the CKM matrix.
Thus $N_d$ and $N_u$ are responsible for
flavor changing neutral currents (FCNC)
involving the scalars $R$ and $I$.

Let us now look back at Table~\ref{master2}.
We see that $\theta=0$,
corresponding to the usual definition of CP,
forces all Yukawa couplings to be real.
This is well known,
but it requires that CP violation arise
spontaneously.
This theory leads to FCNC,
which are constrained by experiment.
This could be cured with further discrete symmetries.
It is known that imposing the absence of FCNC
is inconsistent with GCP for three quark families and
any number of Higgs fields \cite{GCP_NFC}.
Here we will take the view that the scalar masses may be large enough
to suppress FCNC,
and consider the most general couplings.

The case of $\theta=\pi/2$ corresponds to CP2.
As we have seen, this forces
$(\Gamma_{1})_{33} = (\Gamma_{2})_{33} = 0$.
If $\alpha=\beta=\pi/2$ then the $mn$ block vanishes.
Thus the determinant of $D_d$ vanishes,
forcing one quark mass to zero.
This is excluded by experiment.
If $\alpha \neq \pi/2$ and/or $\beta \neq \pi/2$,
then the last column and/or the last row of $D_d$ vanishes,
which leads again to a zero mass quark.
As a result,
\textit{it is impossible to extend CP2 to the quark sector}
in a way consistent with experiment.
This zero mass problem had already been faced in
\cite{NM,NMl},
for a particular extension of CP2.
We note that CP2 could be consistently extended to
the fermion sector provided there were four fermion
families.

Now we come to the most interesting case: consider
$0 < \theta < \pi/2$.
Here $(\Gamma_{1})_{33} = (\Gamma_{2})_{33} = 0$ and,
in order for the $m3$ and $3n$ entries to differ from zero
(otherwise there would be at least a zero mass quark),
we need to have $\alpha = \beta = \theta$. In this
case, the determinants for the $mn$ block in Eq.~(\ref{mndet})
simplify to
\ba
\mbox{Imaginary:}\;\; 256\, (c_{\theta/2})^8 (1 - 2 c_\theta)^2
\label{deti}\\
\mbox{Real:}\;\; 256\, (s_{\theta/2})^8 (1 + 2 c_\theta)^2 .
\label{detr}
\ea
The second determinant never vanishes, which implies that
the $mn$ entries will only have imaginary parts. And the
first determinant, Eq.~(\ref{deti}), only vanishes if
$\theta = \pi/3$. For all other cases, the quark mass
matrices will have zero eigenvalues. Notice that the fact that the
angles $\alpha$, $\beta$ and $\theta$ are constrained to be in the
interval $[0, \pi/2]$ was fundamental in this demonstration.

Therefore, out of the infinite number of GCP symmetries,
only one survives:
the case where one has $\theta = \alpha = \beta = \pi/3$.
For this model the Yukawa matrices have the form:
\ba
\Gamma_1 &=&
\left[
\begin{array}{ccc}
i a_{11} & i a_{12} & a_{13}\\
i a_{12} & - i a_{11} & a_{23}\\
a_{31} & a_{32} & 0\\
\end{array}
\right],
\nonumber\\
\Gamma_2 &=&
\left[
\begin{array}{ccc}
i a_{12} & -i a_{11} & -a_{23}\\
-i a_{11} & -i a_{12} & a_{13}\\
-a_{32} & a_{31} & 0\\
\end{array}
\right].
\label{special_gamma}
\ea
Notice that $\Gamma_1$ has 2 (4) independent imaginary (real) entries and that
$\Gamma_2$ does not involve any new parameter.
Similar parametrizations hold for $\Delta_1$ and $\Delta_2$,
involving 6 new parameters ($a \rightarrow b$).
We stress that this is the only possible extension of GCP into the fermion
sector consistent with the fact that the quarks have non-vanishing masses,
and it leads to a very tightly constrained ``minimal'' model.

Our theory has only 12 Yukawa parameters and two independent vevs.
Indeed, $v_1$ may be made real without loss of generality;
given $v^2$, this fixes $v_2$,
and we only need the phase $\delta$.
In order to perform a numerical fit,
we define the matrices
\ba
H_d &=& \tfrac{v^2}{2} \Gamma_1^H (\Gamma_1^H)^\dagger
= U_{d_L} D_d^2 U_{d_L}^\dagger,
\nonumber\\
H_u &=& \tfrac{v^2}{2} \Delta_1^H (\Delta_1^H)^\dagger
= U_{u_L} D_u^2 U_{u_L}^\dagger.
\ea
The eigenvalues of these matrices give the square of the quark masses,
providing six constraints.
There are four other experimental constraints arising from the four
independent parameters parametrizing the CKM matrix $V$.
We can take these as the eigenvalues of $H_u H_d$,
and also the CP-violating quantity \cite{GCP2}
\ba
J &=& \textrm{Tr} [H_u,H_d]^3 =
 6i (m_t^2 - m_c^2)(m_t^2 - m_u^2)(m_c^2 - m_u^2) \nonumber \\
 &\times &
 (m_b^2 - m_s^2)(m_b^2 - m_d^2)(m_s^2 - m_d^2)
\textrm{Im} \left( V_{us} V_{cb} V_{ub}^\ast V_{cs}^\ast \right).
\label{J}
\ea
Using Eq.~(\ref{special_gamma}) in Eq.~(\ref{J}),
we have shown that $\delta = 0$ implies $J=0$.
Said otherwise,
spontaneous CP violation is required in order to have
CP violation in the CKM matrix,
and this only occurs if we add a soft symmetry
breaking term to the scalar potential. However, the
CP properties of this model are far from obvious,
and will be discussed in the following section.

Using a fast numerical fit,
we have found that setting
$a_{11}=4.6927\times 10^{-6}$, $a_{12}=-5.9799\times 10^{-4}$,
$a_{13}=-2.32\times 10^{-2}$, $a_{23}=-6.6\times 10^{-3}$,
$a_{31}=-8.815\times 10^{-5}$, $a_{32}=5.1193\times 10^{-6}$
in the down-quark sector and
$b_{11}=7.3\times 10^{-3}$, $b_{12}=7.6445\times 10^{-5}$,
$b_{13}=9.578\times 10^{-1}$, $b_{23}=2.325\times 10^{-1}$,
$b_{31}=1.3446\times 10^{-4}$, $b_{32}=5.9491\times 10^{-4}$,
in the up-quark sector,
and $v_1=173.944$, $v_2 \cos{\delta}= -0.8467$, $v_2 \sin{\delta}= -0.9565$,
we obtain
$m_d=0.00298$, $m_s=0.10511$, $m_b=4.19701$, $m_u=0.00200$, $m_c=1.27439$,
$m_t= 171.451$,
for the masses,
and the magnitudes of the CKM matrix elements are
\be
|V| =
\left[
\begin{array}{ccc}
0.97430 & 0.22521 & 0.00339\\
0.22516 & 0.97348 & 0.04039\\
0.00579 & 0.04011 & 0.99918
\end{array}
\right].
\ee
All masses and vevs are in GeV.
The values of $v^2=v_1^2+v_2^2$, the masses, and the magnitudes of the
CKM matrix elements are in very good agreement with the experimental data
\cite{PDG}. One may wonder whether the vevs we used above can be generated
by the scalar potential. As we mentioned earlier, we need to introduce
soft breaking terms in the scalar potential to generate $\delta$, namely
by setting $m_{11}^2 \neq m_{22}^2$ and $\textrm{Re}(m_{12}^2) \neq 0$.
It is easy to find values for the potential's parameters which generate
the desired vevs: they can be reproduced by
$m_{11}^2 = -28791$,
$m_{22}^2 = 5679.5$,
$\textrm{Re}(m_{12}^2) = -167.8012$,
$\textrm{Im}(m_{12}^2) = 0$,
$\lambda_1 = \lambda_2 = 0.9516$,
$\lambda_3 = 0.0176$,
and $\lambda_4 = 0.3643$ (the quadratic parameters are expressed in
units of GeV$^2$).

The numerical fit is quite successful, but the exception is
$|V_{td}|$ which, due to the hierarchical
nature of the CKM matrix elements, becomes much more difficult to
fit~\footnote{Notice that the direct measurement of
$|V_{td}|$ comes from the mixing in the
$B$ system, which occurs in the SM through a
box diagram.
This mixing can receive contributions
involving the second Higgs,
thus altering the value of $|V_{td}|$.}.
A related problem occurs with the CP-violating quantity
$J_{\textrm ckm} = \textrm{Im} \left( V_{us} V_{cb} V_{ub}^\ast V_{cs}^\ast \right)$.
For our choice of parameters,
one obtains $J_{\textrm ckm}= -5.9\times 10^{-8}$,
while the experimental data leads to
$J_{\textrm ckm}= (3.05 \pm 0.20) \times 10^{-5}$.
The reason has to do with the sensitivity of our fit procedure to the
input parameters.
We have fit $Z_{us}$, $Z_{ub}$, $Z_{cs}$, and $Z_{cb}$,
where $Z_{ij}=|V_{ij}|^2$.
These four elements can be used to parametrize completely the CKM matrix
\cite{BL,BL1,BL2}.
In particular \cite{BLS},
\be
4 J_{\textrm ckm}^2 =
4  Z_{us} Z_{ub} Z_{cs} Z_{cb}
 - (1 - Z_{us} - Z_{ub} - Z_{cs} - Z_{cb} + Z_{us} Z_{cb} + Z_{ub} Z_{cs} )^2.
\ee
Using the experimental values for the CKM matrix elements \cite{PDG},
one can show that to first order
\ba
\frac{\Delta J_{\textrm ckm}}{J_{\textrm ckm}} & \sim &
540.95 \frac{\Delta Z_{us}}{Z_{us}}
+ 0.7 \frac{\Delta Z_{ub}}{Z_{ub}}
\nonumber\\
&+& 10065.9 \frac{\Delta Z_{cs}}{Z_{cs}}
+ 18.0 \frac{\Delta Z_{cb}}{Z_{cb}}.
\ea
This shows how extremely sensitive $J_{\textrm ckm}$ is to the
exact value of $V_{cs}$,
explaining why it is easy to perform a fast fit to the latter,
but thus inducing a large error on the former.

\section{\label{sec:spontaneous}A New Type of Spontaneous CP Violation}

We now come to the most remarkable feature of this model, namely its
spontaneous CP violation properties. We have already mentioned that, to
obtain a non-zero value for the Jarlskog invariant $J$, Eq.~\eqref{J},
we need the vevs of the scalar fields to have a relative phase. Usually,
this fact by itself is taken to be a sign of spontaneous CP violation in
the scalar sector. The model we are now discussing behaves, however, in
a very different fashion. Regardless of its possible future experimental
relevance as a viable description of nature, this model serves as a
theoretical example of a new kind of spontaneous CP violation.

First, let us demonstrate
that the lagrangian of this model is explicitly CP-conserving, {\em i.e.}
that it is possible to write it in a field basis where all parameters are
real. We see in the CP3 entry in Table~\ref{master1} that the scalar potential
only has real parameters. As was mentioned above, the soft breaking terms
we introduced are also real, so there are no complex phases in the scalar
sector, prior to SESB.

Now for the Yukawa sector. It might seem, looking at the form of
the Yukawa matrices in Eq.~\eqref{special_gamma}, that one cannot avoid
the presence of complex numbers in the Lagrangian. However, we are free
to choose a different quark basis, by performing transformations of the type
\be
q_L \rightarrow U_L q_L\;,\;n_R \rightarrow U_R n_R,
\ee
with $U(3)$ matrices $U_L$ and $U_R$. This changes
 the Yukawa matrices according to
\be
\Gamma_k \rightarrow U_L^\dagger \Gamma_k U_R.
\ee
By choosing $U_L^\dagger = U_R =
\textrm{diag}(e^{-i \pi/4}, e^{-i \pi/4}, e^{i \pi/4})$,
we remove all factors of $i$ from the $\Gamma$ matrices and are left
with
\ba
\Gamma_1 &\rightarrow&
\left[
\begin{array}{ccc}
a_{11} & a_{12} & a_{13}\\
a_{12} & - a_{11} & a_{23}\\
a_{31} & a_{32} & 0\\
\end{array}
\right],
\nonumber\\
\Gamma_2 &\rightarrow&
\left[
\begin{array}{ccc}
a_{12} & - a_{11} & -a_{23}\\
- a_{11} & - a_{12} & a_{13}\\
-a_{32} & a_{31} & 0\\
\end{array}
\right],
\label{gamma_real}
\ea
where all parameters are real. We stress that the basis choice that
leads to this result involves only the fermion fields, and as such does
not introduce any phases in the scalar potential. An identical fermion
basis choice may be used to render real the Yukawa matrices in the
up-quark and leptonic sectors. Hence, we have proved that it is possible
to find a basis where all parameters in the lagrangian are real -
the model is explicitly CP conserving.

After SESB, with the soft breaking we discussed, we introduce a
complex phase $\delta$ in the theory, from the vevs
$v_1/\sqrt{2}$ and $v_2 e^{i \delta}/\sqrt{2}$.
One might wonder whether $\delta \neq 0$
implies the presence of spontaneous
CP violation in the scalar sector.
This is best investigated with the
basis-invariant quantities developed
by Lavoura and Silva \cite{LS} and by Botella and Silva \cite{BS}.
Their explicit calculation shows that they all vanish -
thus, with the proposed soft breaking of the CP3 symmetry
there is no CP violation in the scalar sector,
whether explicit or spontaneous. A simple way to confirm this is by
changing to the Higgs basis, as given by the transformation
in Eq.~\eqref{eq:HBT}. In that basis, the potential, with the soft
breaking terms already included, becomes
\begin{eqnarray}
V_H
&=&
\bar m_{11}^2 H_1^\dagger H_1 + \bar m_{22}^2 H_2^\dagger H_2
- \left[ \bar m_{12}^2 H_1^\dagger H_2 + \textrm{H.c.} \right]
+
\tfrac{1}{2} \bar \lambda_1 (H_1^\dagger H_1)^2
+ \tfrac{1}{2} \bar \lambda_2 (H_2^\dagger H_2)^2
\nonumber\\[6pt]
&+&
\bar \lambda_3 (H_1^\dagger H_1) (H_2^\dagger H_2)
+ \bar \lambda_4 (H_1^\dagger H_2) (H_2^\dagger H_1)
+
\left[
\tfrac{1}{2} \bar \lambda_5 (H_1^\dagger H_2)^2
+ \bar \lambda_6 (H_1^\dagger H_1) (H_1^\dagger H_2)
\right.
\nonumber\\[6pt]
&+&
\left.
\bar \lambda_7 (H_2^\dagger H_2) (H_1^\dagger H_2)
+ \textrm{H.c.}
\right].
\label{VH2}
\end{eqnarray}
All coefficients are real,
except $\bar m_{12}^2$,
$\bar \lambda_5$,
$\bar \lambda_6$,
and
$\bar \lambda_7$.
The stationarity conditions for the vacuum
impose $2 \bar m_{12}^2 = \bar \lambda_6 v^2$.
Thus,
to study CP we need only the phases of
\begin{eqnarray}
- \bar \lambda_7
=
\bar \lambda_6
&=&
\frac{2}{v^4}\ v_1 v_2 (\lambda_1 - \lambda_3 - \lambda_4)
\sin{\delta}\
\eta,
\nonumber\\
\bar \lambda_5
&=&
-
\frac{1}{v^4}\
(\lambda_1 - \lambda_3 - \lambda_4)\,
\eta^2,
\end{eqnarray}
where
\be
\eta = -i\left( v_1^2 e^{i \delta } +  v_2^2 e^{- i \delta } \right).
\ee
The only phases invariant under a trivial rephasing of $H_2$
and, thus,
possibly signaling CP violation are
$\textrm{Im} (\bar \lambda_6\, \bar \lambda_7^\ast)$,
$\textrm{Im} (\bar \lambda_5^\ast\, \bar \lambda_6^2)$,
and $\textrm{Im} (\bar \lambda_5^\ast\, \bar \lambda_7^2)$.
These are precisely the three quantities introduced by
Lavoura and Silva \cite{LS} to probe all possible sources of CP violation
in the scalar sector in a basis-invariant way.
Since these quantities are all zero,
we conclude that there is no CP violation in the
scalar sector of the theory,
even after SESB. This means, in
particular, that one can find a field basis where there is no mixing between the
CP-even and CP-odd neutral scalar particles.

One may then ask whether the phase $\delta$ is at all
relevant in the model. The answer is yes because, due to the interplay between
the scalar and Yukawa sectors of the theory, there is no choice of field basis
through which one can absorb $\delta$. In fact,
the CP-violating quantity $J$ of Eq.~\eqref{J} is directly proportional
to $\sin\delta$. Notice, too, that as long as $\delta\neq 0$ the FCNC
involving $N_d$ and $N_u$, in Eq.~\eqref{eq:N}, will also, in general,
involve complex phases and as such may well serve as further sources
of CP violation.\footnote{If the masses of the new scalars are large enough,
then these sources of CP violation will have a small impact
on current experiments,
including those currently used to constrain $V$.
Our aim in this section is to highlight a new scenario
for spontaneous CP violation
and not a particular implementation thereof.}
But all possible sources of CP violation vanish if
$\delta = 0$, even though no CP breaking occurs in the scalar sector.

This, then, is a new type of CP violation:
\begin{itemize}
\item It is not like that of the SM, since there CP is explicitly
broken at the lagrangian level;
\item It is not like the CP violation that occurs in Lee-type
mechanisms, since there CP is spontaneously broken in the scalar sector.
\end{itemize}
In the model herein presented, it is the fermion sector which exhibits
the CP violation that arises spontaneously. And the CKM matrix is generated
through a spontaneous breaking of CP, not an explicit one as is the case of
the SM.
Of course, these results were obtained at tree level.
Still,
it is interesting that the scalar sector
does the deed (spontaneously break the symmetry)
but it is the fermion sector which pays the
consequence (providing CP violation).
To the best of our knowledge, this type of CP violation is
unheard of in the literature.

\section{\label{sec:conclusions}Conclusions}

We have shown that,
for the most part,
the generalized CP symmetries of the scalar sector
cannot be extended to the quark sector with
three generations,
while keeping all six quarks massive.
This reduces dramatically the types of available models.
We have found the only exception: the only THDM with GCP which
leads to an acceptable fermion mass spectrum. We have
shown that the model has very few parameters, both in the
scalar and Yukawa sectors,
and that it can give a good fit to the known quark masses and mixings.
Due to the hierarchical nature of the masses and CKM matrix elements,
very precise numerical fits are needed. This model has FCNC which might
have considerable impact on its experimental predictions for
hadron phenomenology. A detailed study of its impact on such
observables is therefore needed, and will be presented
elsewhere~\cite{preparation}.

We have also shown that this model has the peculiar feature that
the scalar sector by itself is both explicitly and spontaneously
CP-conserving,
but that it generates a symmetry breaking which induces CP violation
in the quark sector. This constitutes a new type of spontaneous
CP violation, different from the SM or the usual findings of
the THDM. As such this model, regardless of its experimental
relevance, serves as the first example of a new way of generating CP
violation. It is not unique in that respect: in fact, it is a trivial
exercise to build a three-Higgs doublet model which displays
exactly the same CP properties.

There is a further motivation for studying in detail
models of spontaneous CP violation.
The excellent agreement between the SM
and all CP violating experiments has been taken
as a definitive confirmation of the SM's
CP violation mechanism, {\em i.e.}
the existence of explicitly CP violating
Yukawa couplings.
Models of CP violation,
like the one presented here,
where $J$ could be the dominant source
of CP violation raise the tantalizing possibility that,
contrary to widespread belief,
the current experiments on CP violation
confirm the Cabibbo-Kobayashi-Maskawa
source of CP violation but not its origin in the
explicit CP breaking of the Yukawa interactions.


{\bf Acknowledgments}

We are grateful to L.\ Lavoura for reading and commenting
on the manuscript.
The work of P.M.F. is supported in part by the Portuguese
\textit{Funda\c{c}\~{a}o para a Ci\^{e}ncia e a Tecnologia} (FCT)
under contract PTDC/FIS/70156/2006.
The work of J.P.S. is supported in
part by FCT under contract CFTP-Plurianual (U777).


\begin{thebibliography}{99}

\bibitem{Lee}
T.\ D.\ Lee, ``A Theory of Spontaneous T Violation,''
Phys.\ Rev.\ D \textbf{8}, 1226 (1973),
.
%
\bibitem{GCP1}
T.\ D.\ Lee and G.\ C.\ Wick,
``Space Inversion, Time Reversal, And Other Discrete Symmetries In Local Field Theories,''
Phys.\ Rev.\ \textbf{148}, 1385 (1966).
%
\bibitem{GCP2}
J.\ Bernab\'eu, G.\ C.\ Branco, and M.\ Gronau,
``Cp Restrictions On Quark Mass Matrices,''
Phys.\ Lett.\ \textbf{169B}, 243 (1986).
%
\bibitem{GCP3}
G.\ Ecker, W.\ Grimus, and W.\ Konetschny,
``Quark Mass Matrices In Left-Right Symmetric Gauge Theories,''
Nucl.\ Phys.\ \textbf{B191}, 465 (1981).
%
\bibitem{GCP4}
  G.~Ecker, W.~Grimus and H.~Neufeld,
  ``Spontaneous CP Violation In Left-Right Symmetric Gauge Theories,''
  Nucl.\ Phys.\  B {\bf 247}, 70 (1984).
%
\bibitem{GCP_NFC}
H.~Neufeld, W.~Grimus and G.~Ecker,
   ``Generalized CP invariance, neutral flavor conservation
   and the structure of the mixing matrix'',
  Int.\ J.\ Mod.\ Phys.\  A {\bf 3}, 603 (1988).
%
\bibitem{Ivanov1}
  I.~P.~Ivanov,
   ``Minkowski space structure of the Higgs potential in 2HDM: II. Minima,
  symmetries, and topology,''
  Phys.\ Rev.\  D {\bf 77}, 015017 (2008)
  [arXiv:0710.3490 [hep-ph]].
%
\bibitem{FHS}
  P.~M.~Ferreira, H.~E.~Haber and J.~P.~Silva,
   ``Generalized CP symmetries and special regions of parameter space in the two-Higgs-doublet model,''
  Phys.\ Rev.\  D {\bf 79}, 116004 (2009)
  [arXiv:0902.1537 [hep-ph]].
%
\bibitem{simpleGCP}
  G.~Ecker, W.~Grimus and H.~Neufeld,
  ``A Standard Form For Generalized CP Transformations,''
  J.\ Phys.\ A  {\bf 20}, L807 (1987).
%
\bibitem{details1}
L.~Lavoura,
  ``Models of CP violation exclusively via neutral scalar exchange,''
  Int.\ J.\ Mod.\ Phys.\  A {\bf 9}, 1873 (1994).
%
\bibitem{BS}
F.~J.~Botella and J.~P.~Silva,
  ``Jarlskog - like invariants for theories with scalars and fermions,''
  Phys.\ Rev.\  D {\bf 51}, 3870 (1995)
  [arXiv:hep-ph/9411288].
%
\bibitem{NM}
M.~Maniatis, A.~von Manteuffel and O.~Nachtmann,
   ``A new type of CP symmetry, family replication and fermion mass
  hierarchies,''
  Eur.\ Phys.\ J.\  C {\bf 57}, 739 (2008)
  [arXiv:0711.3760 [hep-ph]].
  %
\bibitem{NMl}
  M.~Maniatis and O.~Nachtmann,
   ``On the phenomenology of a two-Higgs-doublet model with maximal CP symmetry at the LHC,''
  JHEP {\bf 0905}, 028 (2009)
  [arXiv:0901.4341 [hep-ph]].
%
\bibitem{PDG}
  C.~Amsler {\it et al.}  [Particle Data Group],
  ``Review of particle physics,''
  Phys.\ Lett.\  B {\bf 667}, 1 (2008).
%
\bibitem{BL}
  G.~C.~Branco and L.~Lavoura,
  ``Rephasing Invariant Parametrization Of The Quark Mixing Matrix,''
  Phys.\ Lett.\  B {\bf 208}, 123 (1988).
%
\bibitem{BL1}
  L.~Lavoura,
   ``Parametrization of the four generation quark mixing matrix by the moduli of its matrix elements,''
  Phys.\ Rev.\  D {\bf 40}, 2440 (1989).
  %
\bibitem{BL2}
L.~Lavoura,
   ``On The Reconstruction Of The Four Generation Ckm Matrix From The Moduli Of
  Its Matrix Elements: Existence Of Discrete Ambiguities,''
  Phys.\ Lett.\  B {\bf 223}, 97 (1989).
%
\bibitem{BLS}
G.\ C.\ Branco, L.\ Lavoura, and J.\ P.\ Silva,
\textit{CP Violation}\, (Oxford University Press, Oxford, 1999).
%
\bibitem{LS}
L.\ Lavoura and J.\ P.\ Silva,
``Fundamental CP violating quantities in a SU(2) x U(1) model with many Higgs doublets'',
Phys.\ Rev.\ D \textbf{50}, 4619 (1994)
  [arXiv:9404276 [hep-ph]].
%
\bibitem{preparation}
P.\ M.\ Ferreira and J.\ P.\ Silva,
in preparation.
%
\end{thebibliography}
\end{document}